\documentclass[prd,twocolumn,superscriptaddress,showpcs,amsmath,amssymb]{revtex4}

\usepackage{graphicx}
\usepackage{dcolumn}
\usepackage{bm}
\usepackage{xspace}
\usepackage{multirow}
\usepackage{booktabs}
\usepackage{longtable}
\usepackage{lineno}
\usepackage{subfigure}
\usepackage{definitions}

\begin{document}

\title{Measurement of branching ratio and \Bs lifetime in the decay $B_s^0\rightarrow \Jpsi f_0(980)$
 at CDF} 

\affiliation{Institute of Physics, Academia Sinica, Taipei, Taiwan 11529, Republic of China} 
\affiliation{Argonne National Laboratory, Argonne, Illinois 60439, USA} 
\affiliation{University of Athens, 157 71 Athens, Greece} 
\affiliation{Institut de Fisica d'Altes Energies, ICREA, Universitat Autonoma de Barcelona, E-08193, Bellaterra (Barcelona), Spain} 
\affiliation{Baylor University, Waco, Texas 76798, USA} 
\affiliation{Istituto Nazionale di Fisica Nucleare Bologna, $^{aa}$University of Bologna, I-40127 Bologna, Italy} 
\affiliation{University of California, Davis, Davis, California 95616, USA} 
\affiliation{University of California, Los Angeles, Los Angeles, California 90024, USA} 
\affiliation{Instituto de Fisica de Cantabria, CSIC-University of Cantabria, 39005 Santander, Spain} 
\affiliation{Carnegie Mellon University, Pittsburgh, Pennsylvania 15213, USA} 
\affiliation{Enrico Fermi Institute, University of Chicago, Chicago, Illinois 60637, USA}
\affiliation{Comenius University, 842 48 Bratislava, Slovakia; Institute of Experimental Physics, 040 01 Kosice, Slovakia} 
\affiliation{Joint Institute for Nuclear Research, RU-141980 Dubna, Russia} 
\affiliation{Duke University, Durham, North Carolina 27708, USA} 
\affiliation{Fermi National Accelerator Laboratory, Batavia, Illinois 60510, USA} 
\affiliation{University of Florida, Gainesville, Florida 32611, USA} 
\affiliation{Laboratori Nazionali di Frascati, Istituto Nazionale di Fisica Nucleare, I-00044 Frascati, Italy} 
\affiliation{University of Geneva, CH-1211 Geneva 4, Switzerland} 
\affiliation{Glasgow University, Glasgow G12 8QQ, United Kingdom} 
\affiliation{Harvard University, Cambridge, Massachusetts 02138, USA} 
\affiliation{Division of High Energy Physics, Department of Physics, University of Helsinki and Helsinki Institute of Physics, FIN-00014, Helsinki, Finland} 
\affiliation{University of Illinois, Urbana, Illinois 61801, USA} 
\affiliation{The Johns Hopkins University, Baltimore, Maryland 21218, USA} 
\affiliation{Institut f\"{u}r Experimentelle Kernphysik, Karlsruhe Institute of Technology, D-76131 Karlsruhe, Germany} 
\affiliation{Center for High Energy Physics: Kyungpook National University, Daegu 702-701, Korea; Seoul National University, Seoul 151-742, Korea; Sungkyunkwan University, Suwon 440-746, Korea; Korea Institute of Science and Technology Information, Daejeon 305-806, Korea; Chonnam National University, Gwangju 500-757, Korea; Chonbuk National University, Jeonju 561-756, Korea} 
\affiliation{Ernest Orlando Lawrence Berkeley National Laboratory, Berkeley, California 94720, USA} 
\affiliation{University of Liverpool, Liverpool L69 7ZE, United Kingdom} 
\affiliation{University College London, London WC1E 6BT, United Kingdom} 
\affiliation{Centro de Investigaciones Energeticas Medioambientales y Tecnologicas, E-28040 Madrid, Spain} 
\affiliation{Massachusetts Institute of Technology, Cambridge, Massachusetts 02139, USA} 
\affiliation{Institute of Particle Physics: McGill University, Montr\'{e}al, Qu\'{e}bec, Canada H3A~2T8; Simon Fraser University, Burnaby, British Columbia, Canada V5A~1S6; University of Toronto, Toronto, Ontario, Canada M5S~1A7; and TRIUMF, Vancouver, British Columbia, Canada V6T~2A3} 
\affiliation{University of Michigan, Ann Arbor, Michigan 48109, USA} 
\affiliation{Michigan State University, East Lansing, Michigan 48824, USA}
\affiliation{Institution for Theoretical and Experimental Physics, ITEP, Moscow 117259, Russia}
\affiliation{University of New Mexico, Albuquerque, New Mexico 87131, USA} 
\affiliation{Northwestern University, Evanston, Illinois 60208, USA} 
\affiliation{The Ohio State University, Columbus, Ohio 43210, USA} 
\affiliation{Okayama University, Okayama 700-8530, Japan} 
\affiliation{Osaka City University, Osaka 588, Japan} 
\affiliation{University of Oxford, Oxford OX1 3RH, United Kingdom} 
\affiliation{Istituto Nazionale di Fisica Nucleare, Sezione di Padova-Trento, $^{bb}$University of Padova, I-35131 Padova, Italy} 
\affiliation{LPNHE, Universite Pierre et Marie Curie/IN2P3-CNRS, UMR7585, Paris, F-75252 France} 
\affiliation{University of Pennsylvania, Philadelphia, Pennsylvania 19104, USA}
\affiliation{Istituto Nazionale di Fisica Nucleare Pisa, $^{cc}$University of Pisa, $^{dd}$University of Siena and $^{ee}$Scuola Normale Superiore, I-56127 Pisa, Italy} 
\affiliation{University of Pittsburgh, Pittsburgh, Pennsylvania 15260, USA} 
\affiliation{Purdue University, West Lafayette, Indiana 47907, USA} 
\affiliation{University of Rochester, Rochester, New York 14627, USA} 
\affiliation{The Rockefeller University, New York, New York 10065, USA} 
\affiliation{Istituto Nazionale di Fisica Nucleare, Sezione di Roma 1, $^{ff}$Sapienza Universit\`{a} di Roma, I-00185 Roma, Italy} 

\affiliation{Rutgers University, Piscataway, New Jersey 08855, USA} 
\affiliation{Texas A\&M University, College Station, Texas 77843, USA} 
\affiliation{Istituto Nazionale di Fisica Nucleare Trieste/Udine, I-34100 Trieste, $^{gg}$University of Udine, I-33100 Udine, Italy} 
\affiliation{University of Tsukuba, Tsukuba, Ibaraki 305, Japan} 
\affiliation{Tufts University, Medford, Massachusetts 02155, USA} 
\affiliation{University of Virginia, Charlottesville, Virginia 22906, USA}
\affiliation{Waseda University, Tokyo 169, Japan} 
\affiliation{Wayne State University, Detroit, Michigan 48201, USA} 
\affiliation{University of Wisconsin, Madison, Wisconsin 53706, USA} 
\affiliation{Yale University, New Haven, Connecticut 06520, USA} 
\author{T.~Aaltonen}
\affiliation{Division of High Energy Physics, Department of Physics, University of Helsinki and Helsinki Institute of Physics, FIN-00014, Helsinki, Finland}
\author{B.~\'{A}lvarez~Gonz\'{a}lez$^w$}
\affiliation{Instituto de Fisica de Cantabria, CSIC-University of Cantabria, 39005 Santander, Spain}
\author{S.~Amerio}
\affiliation{Istituto Nazionale di Fisica Nucleare, Sezione di Padova-Trento, $^{bb}$University of Padova, I-35131 Padova, Italy} 

\author{D.~Amidei}
\affiliation{University of Michigan, Ann Arbor, Michigan 48109, USA}
\author{A.~Anastassov}
\affiliation{Northwestern University, Evanston, Illinois 60208, USA}
\author{A.~Annovi}
\affiliation{Laboratori Nazionali di Frascati, Istituto Nazionale di Fisica Nucleare, I-00044 Frascati, Italy}
\author{J.~Antos}
\affiliation{Comenius University, 842 48 Bratislava, Slovakia; Institute of Experimental Physics, 040 01 Kosice, Slovakia}
\author{G.~Apollinari}
\affiliation{Fermi National Accelerator Laboratory, Batavia, Illinois 60510, USA}
\author{J.A.~Appel}
\affiliation{Fermi National Accelerator Laboratory, Batavia, Illinois 60510, USA}
\author{A.~Apresyan}
\affiliation{Purdue University, West Lafayette, Indiana 47907, USA}
% ...
\collaboration{CDF Collaboration\footnote{With visitors from $^a$University of MA Amherst,
Amherst, MA 01003, USA,
$^b$Istituto Nazionale di Fisica Nucleare, Sezione di Cagliari, 09042 Monserrato (Cagliari), Italy,
$^c$University of CA Irvine, Irvine, CA  92697, USA,
$^d$University of CA Santa Barbara, Santa Barbara, CA 93106, USA,
$^e$University of CA Santa Cruz, Santa Cruz, CA  95064, USA,
$^f$CERN,CH-1211 Geneva, Switzerland,
$^g$Cornell University, Ithaca, NY  14853, USA, 
$^h$University of Cyprus, Nicosia CY-1678, Cyprus, 
$^i$University College Dublin, Dublin 4, Ireland,
$^j$University of Fukui, Fukui City, Fukui Prefecture, Japan 910-0017,
$^k$Universidad Iberoamericana, Mexico D.F., Mexico,
$^l$Iowa State University, Ames, IA  50011, USA,
$^m$University of Iowa, Iowa City, IA  52242, USA,
$^n$Kinki University, Higashi-Osaka City, Japan 577-8502,
$^o$Kansas State University, Manhattan, KS 66506, USA,
$^p$University of Manchester, Manchester M13 9PL, United Kingdom,
$^q$Queen Mary, University of London, London, E1 4NS, United Kingdom,
$^r$University of Melbourne, Victoria 3010, Australia,
$^s$Muons, Inc., Batavia, IL 60510, USA,
$^t$Nagasaki Institute of Applied Science, Nagasaki, Japan, 
$^u$National Research Nuclear University, Moscow, Russia,
$^v$University of Notre Dame, Notre Dame, IN 46556, USA,
$^w$Universidad de Oviedo, E-33007 Oviedo, Spain, 
$^x$Texas Tech University, Lubbock, TX  79609, USA,
$^y$Universidad Tecnica Federico Santa Maria, 110v Valparaiso, Chile,
$^z$Yarmouk University, Irbid 211-63, Jordan,
$^{hh}$On leave from J.~Stefan Institute, Ljubljana, Slovenia, 
$^{ii}$University of Warwick, Coventry CV4 7AL, United Kingdom,
}}
\noaffiliation

\date{\today}

\begin{abstract}
We present a study of \Bs decays to the \textit{CP}-odd final state $\Jpsi f_0(980)$ 
with $\Jpsi \rightarrow \mu^+ \mu^-$  and $f_0(980)\rightarrow \pi^+\pi^-$. 
Using $p\bar{p}$ collision data with an integrated luminosity
of $3.8$~\invfb collected by the CDF II detector at the Tevatron we
measure a \Bs lifetime of 
$\tau(\Bs\rightarrow\Jpsi f_0(980)) =
1.70_{-0.11}^{+0.12}(\mathrm{stat})\pm 0.03(\mathrm{syst})\,\mathrm{ps}$.
This is the first measurement of the \Bs lifetime in a decay to a \textit{CP} eigenstate and corresponds
in the standard model
to the lifetime of the heavy \Bs eigenstate.
We also measure the product of branching fractions of $\Bs\rightarrow\Jpsi f_0(980)$ and
$f_0(980)\rightarrow \pi^+\pi^-$ relative to the product of
branching fractions of \BsJpsiphi and $\phi\rightarrow
K^+K^-$ to be
$R_{f_0/\phi}=0.257\pm0.020(\mathrm{stat})\pm0.014(\mathrm{syst})$,
which is the most precise determination of this quantity to date.
\end{abstract}


\maketitle

\section{\label{sec:Intro}Introduction}

In the standard model, the mass and flavor eigenstates
of the \Bs meson are not identical. This gives rise to
particle -- anti-particle oscillations \cite{Gay:2001ra}, which proceed in  
the standard model through second order weak interaction processes, and whose
phenomenology depends on the Cabibbo-Kobayashi-Maskawa (CKM)
quark mixing matrix. The time $(t)$ evolution of \Bs mesons
is approximately governed by the Schr\"odinger equation
\begin{equation}
i \frac{d}{dt}
\left(
\begin{array}{c}
| B_s^0(t) \rangle \\ | \bar{B}_s^0 (t) \rangle
\end{array}
\right)
=
\left( \hat{M}^s - \frac{i}{2} \hat{\Gamma}^s \right)
\left(
\begin{array}{c}
| B_s^0(t) \rangle \\ | \bar{B}_s^0 (t) \rangle
\end{array}
\right),
\end{equation}
where $\hat{M}^s$ and $\hat{\Gamma}^s$ are mass and decay
rate symmetric $2\times 2$ matrices. Diagonalization of
$\hat{M}^s - \frac{i}{2} \hat{\Gamma}^s$
leads to mass eigenstates
\begin{eqnarray}
| B_{sL}^0 \rangle&=&p \; | B_s^0 \rangle + q \; | \bar{B}_s^0 \rangle,  \\
| B_{sH}^0 \rangle&=&p \; | B_s^0 \rangle - q \; | \bar{B}_s^0 \rangle,  
\end{eqnarray}
with distinct masses ($M_s^L,\,M_s^H$) and distinct decay rates ($\GammaLs,\,\GammaHs$),
where $p$ and $q$ are complex numbers satisfying $|p|^{2} + |q|^{2}  =  1$.
An important feature of the \Bs system is the non-zero matrix element
$\Gamma_{12}^s$ representing the partial width of \Bs  and $\bar{B}_s^0$ decays to 
common final states which translates into a non-zero decay width
difference $\Delta \Gamma_s$ of the two mass
eigenstates through the relation
\begin{eqnarray}
\Delta \Gamma_s &=& \GammaLs - \GammaHs=  
        2  |\Gamma_{12}^s| \cos  \phi_s, 
\end{eqnarray}
where $\phi_s=\mbox{arg}( -M_{12}^s/\Gamma_{12}^s)$. The
phase $\phi_s$ describes \textit{CP} violation in \Bs mixing. In
the standard model $\phi_s$ is predicted to be
$0.22^\circ\pm0.06^\circ$ \cite{Lenz:2006hd,
Nierste:2011ti}. 
The small value of the phase $\phi_s$ causes the mass and
\textit{CP} eigenstates to coincide to a good approximation. 
Thus the measurement of the lifetime
in a \textit{CP} eigenstate provides directly the lifetime of the
corresponding mass eigenstate. If new physics is present, it
could enhance $\phi_s$ to large values, a scenario which is not
excluded by current experimental constraints. In such a case
the correspondence between mass and \textit{CP} eigenstates does not
hold anymore and the measured lifetime will correspond to the
weighted average of the lifetimes of the two mass eigenstates
with weights dependent on the size of the \textit{CP} violating phase
$\phi_s$~\cite{Dunietz:2000cr}. Thus a measurement of the \Bs lifetime in a final state
which is a \textit{CP} eigenstate provides, in combination with other measurements, 
valuable information on
the decay width difference $\Delta \Gamma_s$ and the \textit{CP} violation in \Bs mixing.

One of the most powerful measurements to constrain a new physics contribution to the
phase $\phi_s$ is the measurement of \textit{CP} violation in the decay \BsJpsiphi with $\phi\rightarrow
K^+K^-$. The decay \BsJpsiphi has a mixture of the \textit{CP}-even and -odd components in the
final state and an angular analysis is needed to separate them~\cite{Dighe:1995pd}. In the standard model,
\textit{CP} violation in the decay \BsJpsiphi is given by
$\beta_s=\arg[({-V_{ts}V_{tb}^{*}})/({V_{cs}V_{cb}^{*}})]$.
New physics effects in \Bs mixing would shift $\phi_s$ and $-2\beta_s$ from the standard model value
by the same amount.
A sufficiently copious \BsJpsifzero signal
with $f_0\rightarrow\pi^+\pi^-$, where $f_0$ stands for $f_0(980)$, and
\Bs flavor identified at production can be used to measure $\beta_s$
without the need of an angular analysis \cite{Stone:2009hd} as $\Jpsi f_0$ is a pure
\textit{CP}-odd final state.
Since the \Bs is a spin 0 particle and the decay products \Jpsi and $f_0$ have
quantum numbers $J^{PC}=1^{--}$ and $0^{++}$, respectively, the final state has
an orbital angular momentum of $L=1$ leading to a \textit{CP} eigenvalue of $(-1)^L=-1$.
Further interest in the decay $\BsJpsifzero$
arises from its possible contribution to an $S$-wave
component in the $\Bs \rightarrow \Jpsi K^+K^-$ decay if the $f_0$ decays to $K^+K^-$.
This contribution could help to resolve  an ambiguity in the
$\Delta \Gamma_s$ and  $\beta_s$ values determined in the \BsJpsiphi analyses.
Because it was neglected in the first tagged \BsJpsiphi 
results~\cite{Aaltonen:2007he,Abazov:2008fj},
each of which showed an approximately 1.5 $\sigma$ deviation from the standard model,
it was argued that the omission may significantly bias
the results~\cite{Stone:2008ak,Stone:2010dp}.
However, using the formalism in Ref.~\cite{Azfar:2010nz}, the latest preliminary CDF 
measurement~\cite{beta_s} has shown that the
$S$-wave interference effect is negligible at the current level of precision.

In Refs.~\cite{Lenz:2006hd,Nierste:2011ti} the decay width
difference in the standard model is predicted to be
$\Delta \Gamma_s^{\rm SM}= (0.087 \pm 0.021)$ ps$^{-1}$
and the ratio of the average \Bs lifetime,
$\tau_s=2/(\GammaLs+\GammaHs)$, to the $B^0$ lifetime,
$\tau_d$, to be $0.996 < \tau_s/\tau_d < 1$.
Using these predictions in the relations
\begin{eqnarray}
\GammaHs&=&\frac{1}{\tauHs}=\Gamma_s-\frac{1}{2}\Delta\Gamma_s, \\
\Gamma_{L}^s&=&\frac{1}{\tauLs}=\Gamma_s+\frac{1}{2}\Delta\Gamma_s,
\end{eqnarray}
where $\Gamma_s=1/\tau_s$,
together with the world average $B^0$ lifetime,
$\tau_d=(1.525\pm0.009)$ ps \cite{Nakamura:2010zzi}, we find
the theoretically-derived values $\tauHs=(1.630\pm0.030)$ ps 
and $\tauLs=(1.427\pm0.023)$ ps.

While no direct measurements of \Bs lifetimes in decays to pure \textit{CP}
eigenstates are available, various experimental results
allow for the determination of the lifetimes of the two mass eigenstates.
Measurements sensitive to these lifetimes are the angular analysis of
\BsJpsiphi decays and the branching fraction of $\Bs\rightarrow
D_s^{(*)+}D_s^{(*)-}$, which can be complemented by
measurements of the \Bs lifetime in flavor specific final
states. The combination of available measurements yields
$\tauHs=(1.544\pm0.041)$ ps and
$\tauLs=(1.407^{+0.028}_{-0.026})$ ps \cite{Asner:2010qj}.
From CDF
measurements we can infer the two lifetimes from the result of
the angular analysis of \BsJpsiphi decays. The latest preliminary result~\cite{beta_s},
that is not yet included in the above average, yields $\tauHs=(1.622\pm0.068)$ ps and
$\tauLs=(1.446\pm0.035)$ ps assuming standard model \textit{CP} violation.

Compared to measurements using \BsJpsiphi decays, lifetime and future \textit{CP} violation 
measurements in the \BsJpsifzero decay suffer from a lower branching fraction. 
Based on a comparison to $D_s^+$ meson decays
Ref.~\cite{Stone:2008ak} makes a prediction for the branching
fraction of \BsJpsifzero decay relative to the \BsJpsiphi decay,
\begin{equation}
R_{f_0/\phi}=\frac{\mathcal{B}(\BsJpsifzero)}{\mathcal{B}(\BsJpsiphi)}
\frac{\mathcal{B}(f_0\rightarrow
\pi^+\pi^-)}{\mathcal{B}(\phi\rightarrow
K^+K^-)},
\end{equation}
to be approximately 0.2.
The CLEO experiment estimates $R_{f_0/\phi}=0.42\pm0.11$
from a measurement of semileptonic $D_s^+$ decays \cite{Ecklund:2009fia}. 
A theoretical prediction based on QCD factorization yields a
range of $R_{f_0/\phi}$ between $0.2$ and $0.5$~\cite{Leitner:2010fq}.
With the world average branching
fraction for the \BsJpsiphi decay of $(1.3\pm0.4)\times 10^{-3}$
and the branching fraction of $f_0\rightarrow\pi^+\pi^-$ in
the region between 0.5--0.8, predictions of 
$\mathcal{B}(\BsJpsifzero)$~\cite{Colangelo:2010bg,Colangelo:2010wg}
translate into a wide range of $R_{f_0/\phi}$ values of approximately 0.1--0.5.

The first experimental search was performed by the Belle
experiment
\cite{Louvot:2010es}.
Their preliminary result did not yield a signal and they
extract an upper limit on the branching fraction of
$\mathcal{B}(\BsJpsifzero)\mathcal{B}(f_0\rightarrow\pi^+\pi^-)<
 1.63\times 10^{-4}\;\mathrm{at\; 90\%\; C.L}$.
Recently the LHCb experiment reported the first observation
of the decay
\BsJpsifzero \cite{Aaij:2011fx} with a relative branching fraction of
$R_{f_0/\phi}=0.252^{+0.046}_{-0.032}(\mathrm{stat})\, ^{+0.027}_{-0.033}(\mathrm{syst})$.
Shortly after the LHCb result was presented, the Belle collaboration announced
their result of an updated analysis using 121.4 \invfb of
$\Upsilon(5S)$ data \cite{Li:2011pg}. They observe
a significant \BsJpsifzero signal and
measure
$\mathcal{B}(\BsJpsifzero)\mathcal{B}(f_0\rightarrow\pi^+\pi^-)=
(1.16^{+0.31}_{-0.19}\, ^{+0.15}_{-0.17}\,
^{+0.26}_{-0.18})\times 10^{-4}$,
where the first uncertainty is statistical, the second
systematic, and the third one is an uncertainty on the
number of produced $B_s^{(*)0}\bar{B}_s^{(*)0}$ pairs. Using their preliminary measurement of the
\BsJpsiphi branching fraction \cite{Louvot:2009xg}, and assuming that the uncertainty on the number of produced
$B_s^{(*)0}\bar{B}_s^{(*)0}$ pairs is fully correlated for the two measurements, this translates into
$R_{f_0/\phi}=0.206^{+0.055}_{-0.034}(\mathrm{stat})\pm0.052(\mathrm{syst})$.
A preliminary measurement of the D0 experiment yields 
$R_{f_0/\phi} = 0.210 \pm 0.032(\mathrm{stat}) \pm 0.036(\mathrm{syst})$~\cite{D0:6152}.

In this paper we present a measurement of the ratio $R_{f_0/\phi}$ of the branching
fraction of the \BsJpsifzero decay relative to the \BsJpsiphi decay
and the first measurement of the \Bs lifetime in a decay to
a pure \textit{CP} eigenstate.  We
use data collected by the CDF II detector from February 2002
until October 2008. The data
correspond to an integrated luminosity of $3.8\:\mbox{fb}^{-1}$.

This paper is organized as follows: In Sec.~\ref{sec:detector} we describe the CDF II detector together with
the online data selection, followed by the candidate selection in
Sec.~\ref{sec:selection}. Section \ref{sec:bf} describes
details of the measurement of the ratio $R_{f_0/\phi}$ of
branching fractions of the \BsJpsifzero decay relative to
the \BsJpsiphi decay while Sec.~\ref{sec:lifetime} discusses
the lifetime measurement. We finish with a short discussion of
the results and conclusions in Sec.~\ref{sec:conclusions}.

\section{CDF II detector and trigger}
\label{sec:detector}

Among the components of the CDF II
detector~\cite{Acosta:2004yw} the tracking and muon
detection systems are most relevant for this analysis.
The tracking system lies within a uniform, axial magnetic field of $1.4$~T
strength. The inner tracking volume hosts 7 layers of double-sided silicon
micro-strip detectors up to a radius of
$28$~cm~\cite{Hill:2004qb}. An additional layer
of single-sided silicon is mounted directly on the beam-pipe
at a radius of $1.5$~cm, providing an excellent resolution
of the impact parameter $d_0$, defined as the distance of
closest approach of the track to the interaction point in
the transverse plane. The silicon tracker provides a pseudorapidity coverage up to
$|\eta|\le 2.0$.
The remainder of the tracking volume up to a radius of
$137$~cm is occupied by an open-cell drift chamber
\cite{Affolder:2003ep}. The drift chamber provides up to 96
measurements along the track with half of them being axial and
other half stereo. Tracks with $|\eta|\le 1.0$ pass the full radial extent of
the drift chamber.
The integrated tracking system achieves a transverse momentum resolution of 
$\sigma(p_T)/p^2_T \approx 0.07\%$ (GeV$/c$)$^{-1}$ and an impact parameter resolution 
of $\sigma(d_0) \approx 35$ \mum for tracks with
a transverse momentum greater than 2 \gevc.
The tracking system is surrounded by electromagnetic and hadronic calorimeters, which
cover the full pseudorapidity range of the tracking system
\cite{Balka:1987ty,Bertolucci:1987zn,Albrow:2001jw,Apollinari:1998bg}.
We detect muons in three sets of multi-wire drift chambers. The central muon detector
has a pseudorapidity coverage of $|\eta|<0.6$ \cite{Ascoli:1987av} and
the calorimeters in front of it provide about 5.5 interaction lengths of material.
The minimum transverse momentum to reach this set of muon chambers is about 1.4 \gevc.
The second set of chambers covers the same range in $\eta$, but is located behind an additional
60 cm of steel absorber, which corresponds to about 3 interaction lengths. It has a higher
transverse momentum threshold of 2 \gevc, but provides a cleaner muon
identification. The third set of muon detectors extends 
the coverage to a region of $0.6<|\eta|<1.0$ and
is shielded by about 6 interaction lengths of material.

A three-level trigger system is used for the online event
selection.  The trigger component most important for this
analysis 
is the extremely fast tracker (XFT) \cite{Thomson:2002xp}, 
which at the first level groups hits from the drift chamber
into tracks in the plane transverse to the beamline.
Candidate events containing $\Jpsi \rightarrow \mu^+ \mu^-$ decays
are selected by a dimuon trigger \cite{Acosta:2004yw}
which requires two tracks of opposite charge found by the XFT 
that match to track segments in the muon
chambers and have a dimuon invariant mass in the range 2.7 to 4.0
GeV/$c^2$.

\section{Reconstruction and candidate selection}
\label{sec:selection}

\subsection{Reconstruction}

In the offline reconstruction we first combine two muon
candidates of opposite charge to form a \Jpsi candidate. 
We consider all tracks that can be matched to a track segment 
in the muon detectors as muon candidates.
The \Jpsi candidate is subject to
a kinematic fit with a vertex constraint. We then
combine the \Jpsi candidate with two other oppositely charged tracks that are
assumed to be pions and have an invariant mass between 0.85 and 1.2
\gevcc to form a \BsJpsifzero candidate. In the final step
a kinematic fit of the \BsJpsifzero candidate is performed. In
this fit we constrain all four tracks to originate from a common vertex,
and the two muons forming the \Jpsi are constrained to have
an invariant mass equal to the world average \Jpsi mass \cite{Nakamura:2010zzi}. In
a similar way we also reconstruct \BsJpsiphi candidates
using pairs of tracks of opposite charge assumed to be kaons and
having an invariant mass between 1.009 and 1.029 \gevcc.
During the reconstruction we place minimal requirements on
the track quality, the quality of the kinematic fit, and the
transverse momentum of the $\Bs$ candidate to ensure high
quality measurements of properties for each candidate.
For the branching fraction measurement we add a requirement
which aims at removing a large fraction of short-lived background.
We require the decay time of the \Bs candidate in its own rest frame, the proper decay time,
to be larger than three times its uncertainty.
This criterion is not imposed in the lifetime analysis since
it would bias the lifetime distribution.
The proper decay time is determined by the expression
\begin{equation}
t=\frac{L_{xy}\cdot m(\Bs)}{c\cdot p_T}
\end{equation}
where $L_{xy}$ is the flight distance projected onto the \Bs momentum in the plane transverse to the
beamline, $p_T$ is the transverse momentum of the given candidate, and
$m(\Bs)$ is the reconstructed mass of the \Bs candidate.
The uncertainty on the proper decay time $t$ is estimated for
each candidate by propagating track parameter and primary
vertex uncertainties into an uncertainty on $L_{xy}$.
The proper decay time resolution is typically of the order of 0.1 ps.

\subsection{Selection}

The selection is performed using a neural network based
on the \textsc{neurobayes} package
\cite{Feindt:2006pm,Feindt:2004aa}.
The neural network combines
several input variables to form a single output variable on
which the selection is performed.
The transformation from the multidimensional space of input variables to
the single output variable is chosen during a training
phase such that it maximizes the separation between signal and
background distributions. For each of the two measurements presented in
this paper we use a specialized neural network.
For the training we need two sets of events with a
known classification of signal or background. For the signal sample
we use simulated events. We generate the kinematic
distributions of $\Bs$ mesons according to the measured $b$-hadron
momentum distribution.  
The decay of the generated \Bs particles into the $\Jpsi f_0$ final state 
is simulated using the \textsc{evtgen} package \cite{Lange:2001uf}.
Each event is passed through the standard CDF II
detector simulation, based on the \textsc{geant3} package
\cite{Brun:1978fy,Gerchtein:2003ba}.
The simulated events are reconstructed with the same
reconstruction software as real data events.
The background sample is taken from data using candidates with the
$\Jpsi\pi^+\pi^-$ invariant mass above the \Bs signal peak,
where only combinatorial background events contribute.
Because the requirement on the proper decay time significance
efficiently suppresses background events in the branching 
ratio measurement, we use an enlarged sideband region
of 5.45 to 5.55 \gevcc in this analysis, compared to an
invariant mass range from  5.45 to 5.475 \gevcc for
the lifetime measurement.

For the branching fraction measurement, the inputs to the
neural network, ordered by the importance of their
contribution to the discrimination power, are the transverse momentum of the $f_0$, the
$\chi^2$ of the kinematic fit of the $\Bs$ candidate using information in
the plane transverse to the beamline, the proper decay time
of the $\Bs$ candidate, the quality of the kinematic fit of the $\Bs$
candidate, the helicity angle of the positive pion, the
transverse momentum of the $\Bs$ candidate, the quality of the kinematic
fit of the two pions with a common vertex constraint, the helicity
angle of the positive muon, and the quality of
the kinematic fit of the two muons with common vertex
constraint.
The helicity angle of the muon (pion) is defined as the angle 
between the three momenta of the muon (pion) and \Bs
candidate measured in the rest frame of the \Jpsi ($f_0$).
For the selection of \BsJpsiphi decays we use
the same neural network without retraining and simply replace
$f_0$ variables by $\phi$ variables and pions by kaons.

For the lifetime measurement we modify the list of inputs
by removing the proper decay time. We also do not use the helicity angles
as they provide almost no additional separation power on the selected sample. 
Since we are not concerned about a precise
efficiency determination for the lifetime measurement, we add the following inputs: the
invariant mass of the two pions, the likelihood based
identification information for muons~\cite{Thesis:Giurgiu}, and the invariant mass
of the muon pair. 
The muon identification is based on the matching of tracks from the tracking system to track segments in
the muon system, energy deposition in the electromagnetic and hadronic calorimeters, and isolation of the
track. The isolation is defined as the transverse momentum carried by the muon candidate over the scalar sum of
transverse momenta of all tracks in a cone of $\Delta R=\sqrt{(\Delta\phi)^2+(\Delta\eta)^2}<0.4$,
where $\Delta\phi$ ($\Delta\eta$) is the difference in azimuthal angle (pseudorapidity) of the muon candidate and the track. 
There is no significant change in the importance
ordering of the inputs. The invariant mass of the pion pair
becomes the second most important input, 
the likelihood based identification of the two muon candidates
is ranked fourth and sixth in the importance list, 
and the muon pair invariant mass is the least important input.

\begin{figure}
\centering
\includegraphics[width=7.0cm]{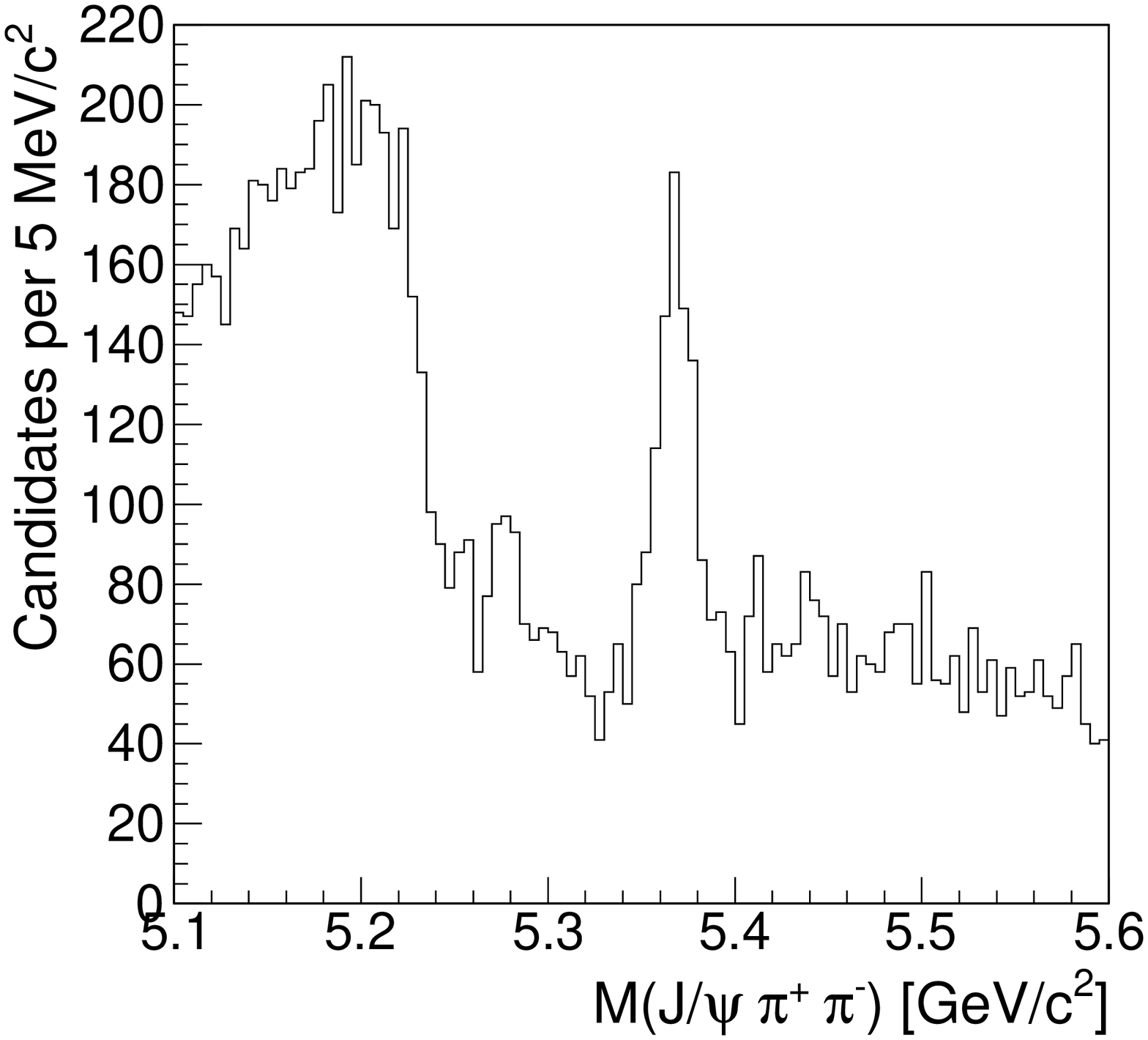}
\caption{The invariant mass distribution of 
$\Bs\rightarrow \Jpsi f_0$ candidates selected for the
branching fraction measurement.}
\label{fig:selectedDataf0}
\end{figure}

\begin{figure}
\centering
\includegraphics[width=7.0cm]{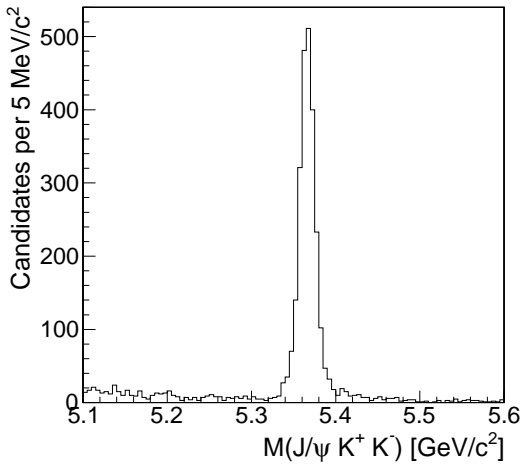}
\caption{The invariant mass distribution of \BsJpsiphi
candidates selected for the branching fraction measurement.}
\label{fig:selectedDataphi}
\end{figure}

For the branching fraction measurement we select the threshold 
on the neural network output by
maximizing $\epsilon/(2.5+\sqrt{N_b})$
\cite{Punzi:2003bu}, where
$\epsilon$ is the reconstruction efficiency for
\BsJpsifzero decays and $N_b$ is the number of background
events estimated from the $\Jpsi\pi^+\pi^-$ mass sideband. The
invariant mass distributions of selected \BsJpsifzero and \BsJpsiphi
candidates are shown in Figs.~\ref{fig:selectedDataf0} and
\ref{fig:selectedDataphi}.
A clear signal at around 5.36 \gevcc is visible in both mass
distributions.


For the lifetime measurement we use simulated experiments 
to determine the optimal neural network output requirement.
We select a value that minimizes the statistical uncertainty of the measured
lifetime. We scan a wide range of neural network output
values and for each requirement we simulate an ensemble of
experiments with a \Bs lifetime of 1.63 ps, where the number of signal and background events as well
as the background distributions are simulated according to data.
For a broad range of selection requirements we observe
the same uncertainty within a few percent. Our final requirement on the network output is
chosen from the central region of this broad range of equivalent options.

\subsection{Physics backgrounds}

\begin{figure}
\centering
\includegraphics[width=7cm]{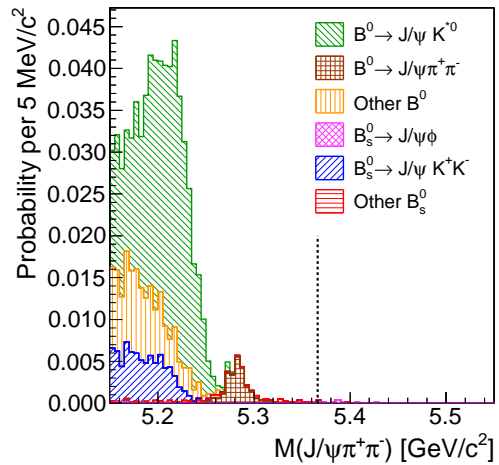}
\caption{(color online) Stacked histogram of physics backgrounds to \BsJpsifzero
derived from simulation using the selection for the branching
fraction measurement.
The vertical line indicates the location of the world average \Bs mass.}
\label{fig:stackedBg}
\end{figure}

We study possible physics backgrounds using simulated events
with all $b$-hadrons produced and
decayed inclusively to final states containing a \Jpsi. For
this study we use the selection from the branching fraction
measurement. While
several physics backgrounds appear in the $\Jpsi\pi^+\pi^-$
mass spectrum, none contributes significantly under
the $\Bs$ peak. The most prominent physics backgrounds are
$B^0\rightarrow \Jpsi K^{*0}$ with
$K^{*0}\rightarrow K^+\pi^-$, where $K^{*0}$ stands for $K^{*}(892)^0$, and $B^0\rightarrow \Jpsi
\pi^+\pi^-$. In the first case the kaon is
mis-reconstructed as a pion and gives rise to a large
fraction of the structure seen below 5.22 \gevcc, while the second one is
correctly reconstructed and produces the narrow peak at approximately 5.28 \gevcc.
Another possible physics type of background would consist of properly reconstructed $B^+$
combined with a random track. 
This type of background would contribute only to higher masses with a threshold above the \Bs signal.
As we do not find evidence of such background in Ref.~\cite{Aaltonen:2011sy} which is more sensitive
we conclude that this kind of background is also negligible here.
The stacked histogram of physics backgrounds derived from
simulation is shown in Fig.~\ref{fig:stackedBg}. From this study
we conclude that the main physics background that has to be included as
a separate component in a fit to the mass spectrum above 5.26 \gevcc stems from decays of
$B^0\rightarrow \Jpsi\pi^+\pi^-$. It is
properly reconstructed and therefore simple to parametrize.
All other physics backgrounds are negligible.

\section{Branching fraction measurement}
\label{sec:bf}

In this Section we describe details of the branching
fraction measurement. These involve the maximum likelihood
fit to extract the number of signal events, the efficiency
estimation, and the systematic uncertainties. We conclude this Section
with the result for the ratio $R_{f_0/\phi}$ of branching fractions
between \BsJpsifzero and \BsJpsiphi decays.

\subsection{Fit description}
\label{sec:BFfit}

We use an unbinned extended maximum
likelihood fit of the invariant mass to extract the number
of $\Bs$ decays in our samples. 
In order to avoid the need for modeling most of the physics background,
we restrict the fit to the mass range from 5.26 \gevcc to 5.5
\gevcc. The likelihood is 
\begin{eqnarray}
\mathcal{L}=\prod_{i=1}^{N}&&
\left[N_s\cdot P_s(m_i)+ N_{cb}\cdot P_{cb}(m_i)+\right. \nonumber \\
&&\left. f_{pb}\cdot N_s\cdot P_{pb}(m_i) + 
 N_{B^0}\cdot P_{B^0}(m_i)\right]\cdot \nonumber \\
&&e^{-(N_s+N_{cb}+N_s\cdot f_{pb}+N_{B^0})},
\end{eqnarray}
where $m_i$ is the invariant mass of the $i$-th candidate and
$N$ is the total number of candidates in the sample. 
The fit components are denoted by the subscripts $s$ for signal, 
$cb$ for combinatorial background, $pb$ for physics 
background, and $B^0$ for $B^0\rightarrow \Jpsi\pi^+\pi^-$ background.
The yields of the components are given by $N_s$, $N_{cb}$, $N_s\cdot f_{pb}$, and $N_{B^0}$,
and their probability density functions (PDFs) by $P_{s,cb,pb,B^0}(m_i)$, respectively.
The physics background yield is parametrized relative to the signal yield 
via the ratio $f_{pb}$ to allow constraining it by other measurements in the
\BsJpsiphi fit.

The signal PDF $P_s(m_i)$ is
parametrized by a sum of two Gaussian functions with
a common mean. The relative size
of the two Gaussians and their widths
are determined from simulated events. Approximately 82\% of the
\BsJpsifzero decays are contained in a narrower Gaussian with
width of
9.4 \mevcc. The broader Gaussian has width of
18.4
\mevcc. In the case of \BsJpsiphi, the narrow Gaussian with a width
of
7.2 \mevcc accounts for 79\% of the signal, with the
rest of the events having a width of 13.3 \mevcc. To take into
account possible differences between simulation and data, we
multiply all widths by a scaling parameter $S_m$.
Because of kinematic differences between $f_0\rightarrow\pi^+\pi^-$ and
$\phi\rightarrow K^+K^- $ we use independent scale factors for both modes.
In the fits all parameters of the PDF are fixed except for 
the scaling parameter $S_m$. In addition the mean of the Gaussians 
is allowed to float in the $\Jpsi K^+K^-$ fit. Doing so we obtain a value that
is consistent with the world average $\Bs$ mass
\cite{Nakamura:2010zzi}.
For the $\Jpsi\pi^+\pi^-$ fit we fix
the position of the signal to the value determined in the
fit to the $\Jpsi K^+K^-$ candidates.

The combinatorial background PDF $P_{cb}(m_i)$ is parametrized
using a linear function. In both fits we leave
its slope floating. In each of the two fits there is one
physics background. In the case of the
$\Jpsi\pi^+\pi^-$ spectrum, the physics background describes properly
reconstructed $B^0\rightarrow\Jpsi\pi^+\pi^-$ decays using
a shape identical to the $\Bs$ signal and position fixed to
the world
average $B^0$ mass \cite{Nakamura:2010zzi}. The number of
$B^0$ events
$N_{B^0}$ is left free in the fit.
For the $\Jpsi K^+K^-$ fit, we have a contribution from
\BzeroJpsiKS decays where the pion from the $K^{*0}$ decay
is mis-reconstructed as a kaon. This contribution peaks at a mass
of approximately 5.36 \gevcc with an asymmetric tail
towards larger masses. The shape itself is parametrized by
a
sum of a Gaussian function and an exponential function convolved
with a Gaussian. The parameters are derived from simulated
\BzeroJpsiKS events. The normalization of this
component relative to the signal is fixed to $f_{pb}=(3.04\pm 0.99)\times 10^{-2}$, which is
derived from the CDF Run I measurement of the ratio of cross
section times branching fraction for \BsJpsiphi and
\BzeroJpsiKS decays \cite{Abe:1996kc}, the world average
branching
fractions for $\phi$ and $K^{*0}$
\cite{Nakamura:2010zzi}, and the ratio of
reconstruction efficiencies obtained from simulation.

The fit determines a yield of $502 \pm 37$ \BsJpsifzero events
and $2302 \pm 49$ \BsJpsiphi events, where the uncertainties are
statistical only.
The number of $B^0$ background events
in the $\Jpsi\pi^+\pi^-$ fit is
$160\pm30$.

\subsection{Efficiency}

To extract the ratio of branching fractions we need
to
estimate the relative efficiency for reconstruction of
$\BsJpsifzero$ with $f_0\rightarrow\pi^+\pi^-$ and
$\BsJpsiphi$ with $\phi\rightarrow K^+K^-$ decays,
$\epsilon_{rel}=\epsilon(\BsJpsiphi)/\epsilon(\BsJpsifzero)$.
We estimate the efficiency using simulated events in which we generate
a single $\Bs$ meson per event. The \Bs meson then decays with
equal
probabilities to \BsJpsifzero or \BsJpsiphi final states with
exclusive $\Jpsi\rightarrow \mu^+\mu^-$, $\phi\rightarrow
K^+K^-$, and $f_0\rightarrow\pi^+\pi^-$. Generated
events are then processed through a detailed detector
simulation and the offline reconstruction
software used to reconstruct data. 
In both cases angular and decay time distributions are generated
assuming no \textit{CP} violation and parameters taken
from the preliminary result of the
angular distributions analysis \cite{beta_s}:
$\tau = 1.529 \pm 0.028 \; \mathrm{ps}$, 
$\Delta\Gamma = 0.075 \pm 0.036\; \mathrm{ps}^{-1}$,
$|A_0|^2 = 0.524 \pm 0.020$, 
and $|A_{||}|^2 = 0.231 \pm 0.021. $
As a strong phase between $A_0$ and $A_{||}$ is not measured
we
use the world average value from \BzeroJpsiKS decays of
$\phi_{||}
= -2.86 \pm 0.11$ \cite{Nakamura:2010zzi} as a reasonable
approximation \cite{Gronau:2008hb}. An additional
peculiarity of the \BsJpsifzero decay is the unusual mass
shape of
the $f_0$ meson. It is modeled using a Flatt\'e
distribution
\cite{Flatte:1976xu} with input parameters measured by the
BES experiment \cite{Ablikim:2004wn} to be
$m_0 = 965\pm8\pm6\;\mevcc$, 
$g_\pi = 165\pm10\pm15\;\mevcc$, 
and $g_K/g_\pi = 4.21\pm0.25\pm0.21$,
where the errors are statistical and systematic, respectively.
The $\phi$ meson mass distribution is modeled using a relativistic Breit-Wigner distribution with
world average values for its parameters \cite{Nakamura:2010zzi}. 
With these inputs to the simulation we find
$\epsilon_{rel}=1.178$, which accounts for the $\phi$ and $f_0$ mass window
selection requirements.

\subsection{Systematic uncertainties}
\label{sec:BFsystematics}

We investigate several sources of systematic uncertainties.
They can be broadly
separated into two classes: one
dealing with assumptions made in the fits that may affect
yields, and the other 
related to assumptions in the efficiency estimation. In the
first class we estimate
uncertainties by refitting data with a modified assumption
and taking the difference with respect to
the original value as an uncertainty. For the second class
we recalculate the efficiency with
a modified assumption and take the difference with respect
to the default
efficiency as an uncertainty
unless specified otherwise. The summary of assigned
uncertainties is given in Table \ref{tab:BFsystematicsSummary}.

For the yield of \BsJpsiphi we investigate the effect of the
assumption on the combinatorial background
shape, the limited knowledge of mis-reconstructed
\BzeroJpsiKS decays and the shape of the
signal PDF. The uncertainty due to the shape of
combinatorial background is estimated by
changing from the first order polynomial to a constant or a
second order polynomial. For
the physics background we vary the normalization of the
component in the fit and use
an alternative shape determined by varying the momentum
distribution and the decay
amplitudes of \BzeroJpsiKS in simulation. Finally, to
estimate the effect of the signal
PDF parametrization we use an alternative model with a single
Gaussian rather than
two Gaussian functions and an alternative shape from simulation,
where we vary the momentum distribution
of the produced $\Bs$ mesons and the decay amplitudes of
the \BsJpsiphi decay.

To estimate the uncertainty on the \BsJpsifzero yield we
follow a procedure similar to that for \BsJpsiphi
and conservatively treat the systematic effects as independent
between the two modes in the calculation of $R_{f_0/\phi}$.
For the sensitivity to the parametrization of
the combinatorial background
we switch to a second order polynomial or a constant as
alternative parametrization.
For the shape of the signal PDF we use
two alternatives, one with a single Gaussian function
instead of two and another one
with two Gaussians, but varying the momentum distribution in
simulation.
We also vary the position of the \Bs signal 
within the uncertainty determined in the $\Jpsi K^+K^-$ fit.

The systematic uncertainty on the relative efficiency stems
from the statistics of
simulation, an imperfect knowledge of the momentum distribution,
physics parameters of
decays like lifetimes or decay amplitudes, and differences in
the efficiencies of the online selection of
events. To estimate the effect of the imperfect knowledge of
the momentum
distribution we vary the momentum distribution of \Bs mesons in the simulation.
The physics parameters entering the simulation are grouped into three
categories, those defining
the $f_0$ mass shape, the ones determining decay
amplitudes in \BsJpsiphi decays,
and those affecting the lifetimes of the two $\Bs$ mass
eigenstates. In the first two
cases we vary each parameter independently and add all changes 
in the efficiency in quadrature.
For the last case we vary the mean lifetime $\tau$ and the
decay width difference
$\Delta\Gamma$ simultaneously and take the largest variation
as the uncertainty.
We add the uncertainty from the third class in quadrature
with all others to obtain the uncertainty due to the parameters
describing the particle decays. The last
effect deals with how events are selected during data
taking. The CDF
trigger has several different sets of requirements for
the selection of events. The ones used in this
analysis can be broadly sorted
into three classes depending on momentum thresholds and
which subdetectors
detected muons. 
The fraction of events for each different class varies
depending on the instantaneous luminosity, which is not simulated.
To estimate the size of a possible effect we calculate
the efficiency for each class
separately and take half of the largest difference as
the uncertainty.

To obtain the total uncertainty we add all partial
uncertainties in quadrature. In
total we assigned $49$ events (2.1\%) as the systematic uncertainty
on the \BsJpsiphi
yield, $18$ events (3.6\%) on the \BsJpsifzero yield, and 0.040 (3.4\%) on
the relative efficiency $\epsilon_{rel}$.
A summary of the systematic uncertainties in the branching 
ratio is provided in Table~\ref{tab:BFsystematicsSummary}.

\begin{table}
\centering
\caption{The summary of assigned systematic uncertainties
for the branching fraction measurement.}
\begin{tabular}{lccc} \hline\hline
Source & $\Jpsi\phi$ yield& $\Jpsi f_0$ yield& $\epsilon_{rel}$ \\
\hline
Combinatorial bckg. & 34 & 16 & $-$ \\
Physics bckg. & 13 & $-$ & $-$ \\
Mass resolution & 32 & 7.9 & $-$ \\
\Bs mass & $-$ & 0.1 & $-$ \\ \hline
Total & 49 & 18  &  $-$ \\ \hline
MC statistics & $-$ & $-$ & 0.012 \\
Momentum distribution & $-$ & $-$ & 0.011 \\
Decay parameters & $-$ & $-$ & 0.033  \\
Trigger composition & $-$ & $-$ & 0.016  \\ \hline
Total & $-$ & $-$ & 0.040 \\ \hline \hline
\end{tabular}
\label{tab:BFsystematicsSummary}
\end{table}

\subsection{Branching fraction result}

\begin{figure}
\centering
\includegraphics[width=7.5cm]{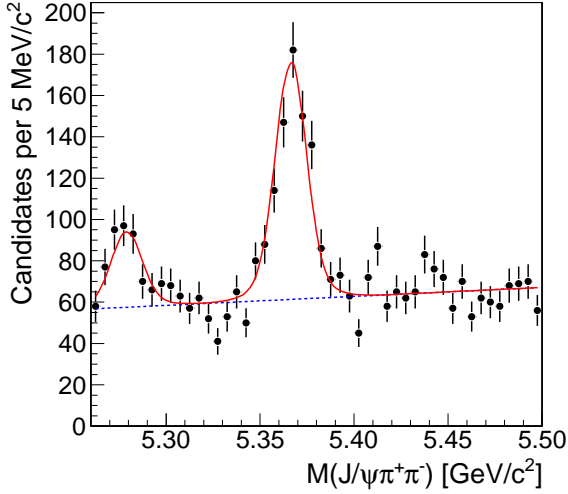}
\caption{(color online) Projection of the fit of the \BsJpsifzero decay mode. 
The dashed line (blue) shows the contribution from combinatorial background.}
\label{fig:BFsignalFit}
\end{figure}
\begin{figure}
\centering
\includegraphics[width=7.5cm]{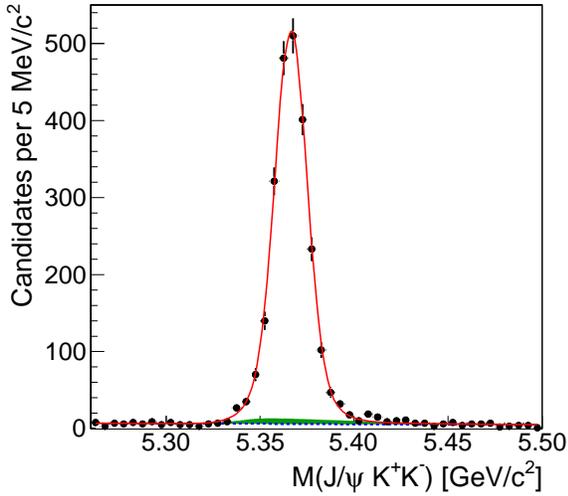}
\caption{(color online) Projection of the fit of the \BsJpsiphi decay mode.
The dashed line (blue) and filled area (green) show the contributions 
from combinatorial background and $B^0\rightarrow \Jpsi K^{*0}$, respectively.}
\label{fig:BFnormalizationFit}
\end{figure}

From the fit we find $502\pm37(\mathrm{stat})\pm18(\mathrm{syst})$ \BsJpsifzero
signal events
and $2302\pm49(\mathrm{stat})\pm49(\mathrm{syst})$ \BsJpsiphi events.
The projections of the fits for \BsJpsifzero and \BsJpsiphi are shown in
Fig.~\ref{fig:BFsignalFit} and
Fig.~\ref{fig:BFnormalizationFit}, respectively.

\begin{figure}
\centering
\includegraphics[width=7cm]{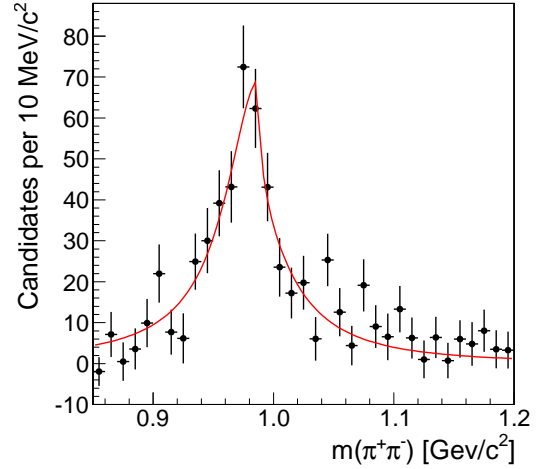}
\caption{The dipion invariant mass distribution after
sideband subtraction with
fit projection overlaid. The fit uses a Flatt\'e distribution
with all parameters
floating.}
\label{fig:f0fit}
\end{figure}
\begin{figure}
\centering
\includegraphics[width=7cm]{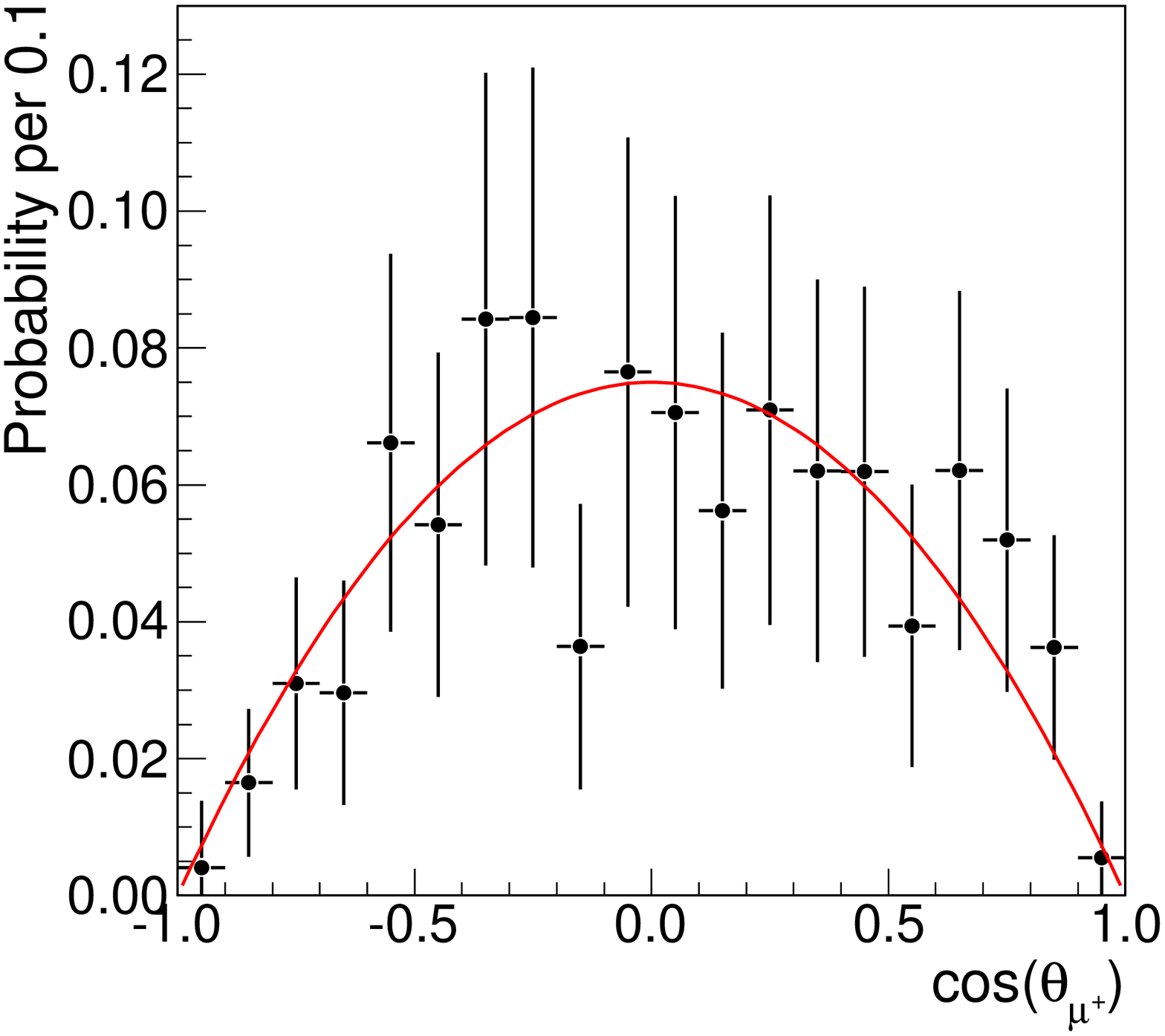}
\caption{Normalized helicity angle distribution for the positive muon
corrected for relative efficiency.
The line shows the expectation for a \BsJpsifzero decay. 
} 
\label{fig:helicityAngles1}
\end{figure}
\begin{figure}
\centering
\includegraphics[width=7cm]{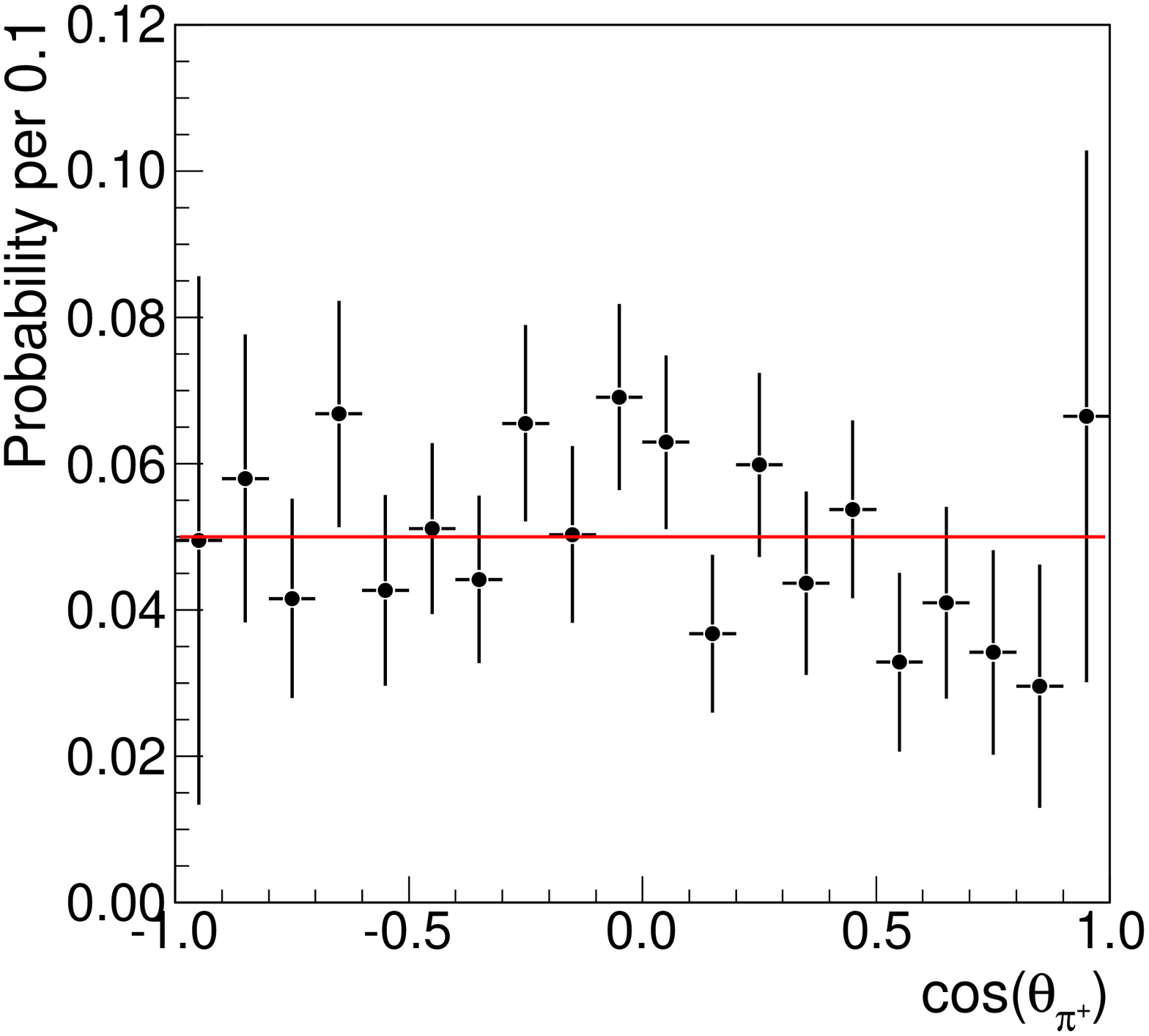}
\caption{Normalized helicity angle distribution for the positive pion
corrected for relative efficiency.
The line shows the expectation for a \BsJpsifzero decay. 
}
\label{fig:helicityAngles2}
\end{figure}

In order to check our interpretation of the signal in
the $\Jpsi\pi^+\pi^-$
distribution being due to the \BsJpsifzero decays we show the
invariant
mass distribution of the pions for \Bs signal data in
Fig.~\ref{fig:f0fit}. To obtain the distribution of \Bs
signal we fit the $\Jpsi\pi^+\pi^-$ mass distribution in the
range 5.26 to 5.45 \gevcc for each bin in $\pi^+\pi^-$ mass
and report the \Bs signal yield as a function of $\pi^+\pi^-$ mass.
We fit the dipion mass distribution using the Flatt\'e
parametrization. The fit probability is 23.4\% and the obtained parameters,
$m_0=989.6\pm9.9(\mathrm{stat})$ \mevcc,
$g_\pi=141\pm19(\mathrm{stat})$  \mevcc, and
$g_K/g_\pi=2.3\pm1.3(\mathrm{stat})$,
are in reasonable agreement with the ones
measured by the BES collaboration \cite{Ablikim:2004wn}.
In Figs.~\ref{fig:helicityAngles1} and
\ref{fig:helicityAngles2} we show the positive muon and pion
helicity angle distributions, obtained in an analogous way to the invariant mass distribution of pion pairs. Those are corrected for
relative efficiencies in the different helicity bins and compared to the
theoretical expectation for a \BsJpsifzero signal.
We use a $\chi^2$ test to evaluate the agreement between 
data and theoretical expectation.
For the distribution of $\cos(\theta_{\mu^+})$
we obtain $\chi^2/\mathrm{ndf}=7.9/20$, which corresponds to 99\%
probability.
Similarly for $\cos(\theta_{\pi^+})$ the $\chi^2/\mathrm{ndf}$ is $15/20$,
giving 78\% probability.
Since the dipion mass as well as the angular distributions are
consistent
with expectations, we interpret our signal as coming solely
from the \BsJpsifzero
decays. On the other hand, as we use a dipion mass window from
0.85 to 1.2 \gevcc, we cannot exclude contributions from other
higher mass states to our
signal with present statistics.

Finally, we obtain the ratio of branching fractions
\begin{eqnarray}
R_{f_0/\phi}&=&\frac{\mathcal{B}(\BsJpsifzero)}{\mathcal{B}(\BsJpsiphi)} 
\frac{\mathcal{B}(f_0\rightarrow
\pi^+\pi^-)}{\mathcal{B}(\phi\rightarrow 
K^+K^-)}= \nonumber \\ && 0.257\pm0.020(\mathrm{stat})\pm0.014(\mathrm{syst}),
\end{eqnarray}
where corrections for events with an $f_0$ or $\phi$ mass outside the ranges
selected in this analysis are taken into account.

\section{Lifetime measurement}
\label{sec:lifetime}

In this Section we discuss the details of the lifetime measurement. 
We describe the maximum likelihood fit, estimate the systematic uncertainties,
and present the result of the lifetime measurement.

\subsection{Fit description}
\label{sec:LFfit}

To extract the \Bs lifetime we use a maximum likelihood fit. The fit
uses three variables: the invariant mass $m_i$, the
decay
time $t_i$, and the decay time uncertainty $\sigma_{ti}$ of
each candidate. 
To exclude $B^0\rightarrow \Jpsi\pi^+\pi^-$ decays we use only candidates with an invariant
mass greater than 5.3 \gevcc in the fit.

The components in the fit are \Bs signal and combinatorial background.
The likelihood function has the form
\begin{eqnarray}
\mathcal{L}=\prod_{i=1}^{N}&&
\left[f_s\cdot P_s(m_i,t_i,\sigma_{ti})+\right. \nonumber \\
&&\left.(1-f_s)\cdot P_{cb}(m_i,t_i,\sigma_{ti})\right].
\label{eqn:Llifetime}
\end{eqnarray}
The parameter $f_s$ denotes the fraction of signal
\BsJpsifzero decays and $P_s$ and $P_{cb}$ the probability
density function of signal and combinatorial background, respectively.
To enhance the signal-to-background ratio in the selected sample,
we use only \Bs candidates with decay times
larger than $0.2$ mm$/c = 0.67$ ps. This requirement 
suppresses background by a factor of 40 and reduces the
prompt background component to a negligible level 
while keeping about two thirds of the signal events.

The \Bs signal mass PDF is parametrized as for
the branching ratio measurement. The PDF in
decay time is parametrized with an exponential function
convolved with a Gaussian resolution function. The width of
the Gaussian is given by the candidate-specific estimated decay
time uncertainty  $\sigma_{ti}$ scaled by a common factor
$S_{t}$ which accounts for possible discrepancies between
estimated and actual resolutions. 
The scaling factor $S_t$ is determined in a fit to data 
dominated by prompt background, selected by requiring $0<t<0.3$ ps.
In the final fit, $S_t$ is a
free parameter with a Gaussian constraint 
included as additional factor in the likelihood in Eq.~(\ref{eqn:Llifetime}). The PDF in
decay time uncertainty is parametrized by an empirical function.
We use a log-normal distribution 
with parameters $\mu$, $\theta$, and $\sigma$ defined as
\begin{equation}
\mathcal{D}(\sigma_{ti}|\mu,\theta,\sigma) = 
\frac{1}{\sqrt{2\pi}\sigma(\sigma_{ti}-\mu)}
\cdotp
e^{-\frac{(\mathrm{ln}(\sigma_{ti}-\mu)-\theta)^2}{2\sigma^2}}
\end{equation}
for $\sigma_{ti}>\mu$ and zero otherwise.
Given the rather small statistics of the \Bs signal we derive
the parameters using simulated \BsJpsifzero events and Gaussian constrain
the values in the fit to data.
The widths of the Gaussian
constraints are chosen to cover possible differences between
simulation and data.

The combinatorial background is described by two components,
a long-lived part for the background
from $b$-hadron decays and a short-lived part for
the tail from mis-reconstructed prompt events.
The mass PDF is common to both components and parametrized by
a linear function.
The decay time PDF of each component is described by an exponential
convolved with the same resolution function as used for signal.
Both decay time uncertainty PDFs are again
modeled using log-normal distributions.
The parameters of each log normal distribution are independent of 
the distribution of the \Bs signal.

All parameters of the combinatorial background are determined from the fit.
The yield, the mass resolution scale factor, and the
lifetime of the \Bs signal are also left to float freely. The decay time
uncertainty parameters of the signal  and the resolution scale parameter
are Gaussian constrained.
Using an ensemble of simulated experiments we verify within 1\%
that the fit is unbiased and returns proper uncertainties.

\subsection{Systematic uncertainties}
\label{sec:LFsystematics}

We investigate several possible sources of systematic uncertainties.
These are broadly separable into two classes: the first dealing with 
the parametrization of the PDFs and the second with possible biases 
in the selection or reconstruction.

We first investigate our assumption of the mass shape of combinatorial background. 
We determine the relative change of the \Bs lifetime between a fit with a first
and a third order polynomial background mass model. For fits in different invariant mass ranges,
we find an average difference of 0.010 ps,
which we assign as the systematic uncertainty. 
The systematic uncertainty assigned to the signal mass shape 
has contributions from the limited knowledge of the mean position 
and from the assumed shape parametrization.
Both effects are evaluated in the same way as for the branching ratio
measurement and yield a systematic uncertainty of 0.005 ps.
There are two assumptions made for the decay time PDFs; one is the
resolution scale factor, $S_t$, which is known only
with limited precision and the other is the shape
of the combinatorial background. The uncertainty of the scale
$S_t$ is included directly in the statistical uncertainty
of the fit as the parameter is allowed to vary within a Gaussian
constraint. To quantify the size of the contribution, we repeat
the fit with $S_t$ fixed to its central value and find the 
quadratic difference in uncertainty to the original fit 
to be 0.005 ps.
To estimate the effect of
the assumed decay time PDF of combinatorial background, we
employ an alternative fit method which does not need a decay time
parametrization of the background. We split the data into
20 decay time bins and simultaneously fit the 
invariant mass distributions with independent parameters
for the background in each bin. 
The signal yield per bin is given by the total signal yield times the integral of the signal 
decay time PDF over the time bin, where the same PDF parametrization
as in the default fit is used.
The difference in the fit results is taken as a measure of the
systematic uncertainty due to the background decay time PDF.
To avoid possible statistical fluctuations in this estimate we repeat the
comparison for different selection requirements and assign
the average difference of 0.021 ps as systematic uncertainty.
The third kind of systematic effect addresses
the uncertainty of the $\sigma_t$ PDFs. The main
effect is the distribution for
signal derived from simulated events. The uncertainty is
already included in the statistical error since the parameters
are Gaussian constrained in the fit. The contribution due to modeling 
of the decay time uncertainty distribution, estimated from a
comparison of fit results with fixed and constrained parameters, is 0.015 ps.

For the second class, we verify that our candidate selection
does not introduce any significant bias. A bias in the mass
distribution could artificially enhance or decrease the amount of
signal candidates while a bias in decay time could directly
affect the extracted lifetime. We verify on a background-enriched
sample selected by requiring $t < 0.01$ cm$/c$ that no artificial peak
is observed for any neural network output requirement.
With a high statistics sample of simulated events we check that
the selection does not bias the fitted lifetime.
A possible lifetime bias introduced by the trigger has been studied in
a previous CDF analysis \cite{Aaltonen:2010pj} and is negligible in our
measurement.
Finally the alignment of the tracking detectors
is known only with finite precision. Previous measurements
found that the uncertainty on the lifetime due to a possible
misalignment is 0.007 ps \cite{Aaltonen:2010pj}.

All the contributions are added in quadrature
and yield a total systematic error on the lifetime of 0.03 ps (1.5\%).
A summary of the systematic uncertainties on the lifetime
is provided in Table~\ref{tab:LFsystematicsSummary}.
\begin{table}
\centering
\caption{Summary of assigned systematic uncertainties for the lifetime measurement.
The uncertainties in parentheses are included in the 
statistical uncertainty via Gaussian constraints in the fit.}
\begin{tabular}{lc} \hline\hline
Source & Uncertainty [ps] \\ \hline
Background mass model & 0.010 \\
Signal mass model & 0.005 \\
Decay time uncertainty scale & ($0.005$) \\
Background decay time model & 0.021 \\
Decay time uncertainty model & (0.015) \\
SVX alignment & 0.007 \\ \hline
Total & 0.03 \\ \hline\hline 
\end{tabular}
\label{tab:LFsystematicsSummary}
\end{table}

\subsection{Lifetime result}
\label{sec:LFresults}

\begin{figure}
\begin{center}
\includegraphics[width=7.5cm]{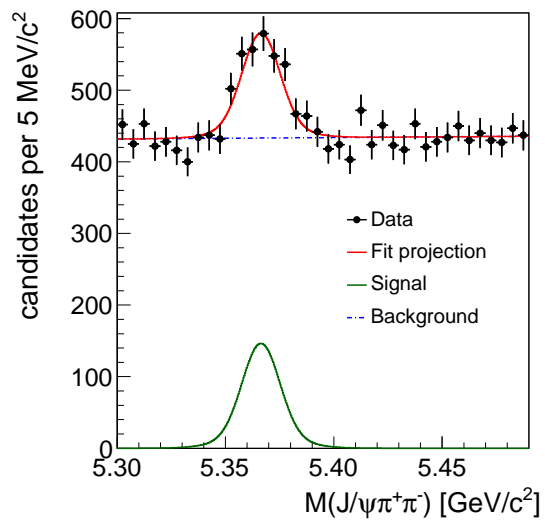}
\caption{(color online) Invariant mass distribution
with fit projection overlaid. 
}
\label{fig:defaultfit1}
\end{center}
\end{figure}
\begin{figure}
\begin{center}
\includegraphics[width=7.5cm]{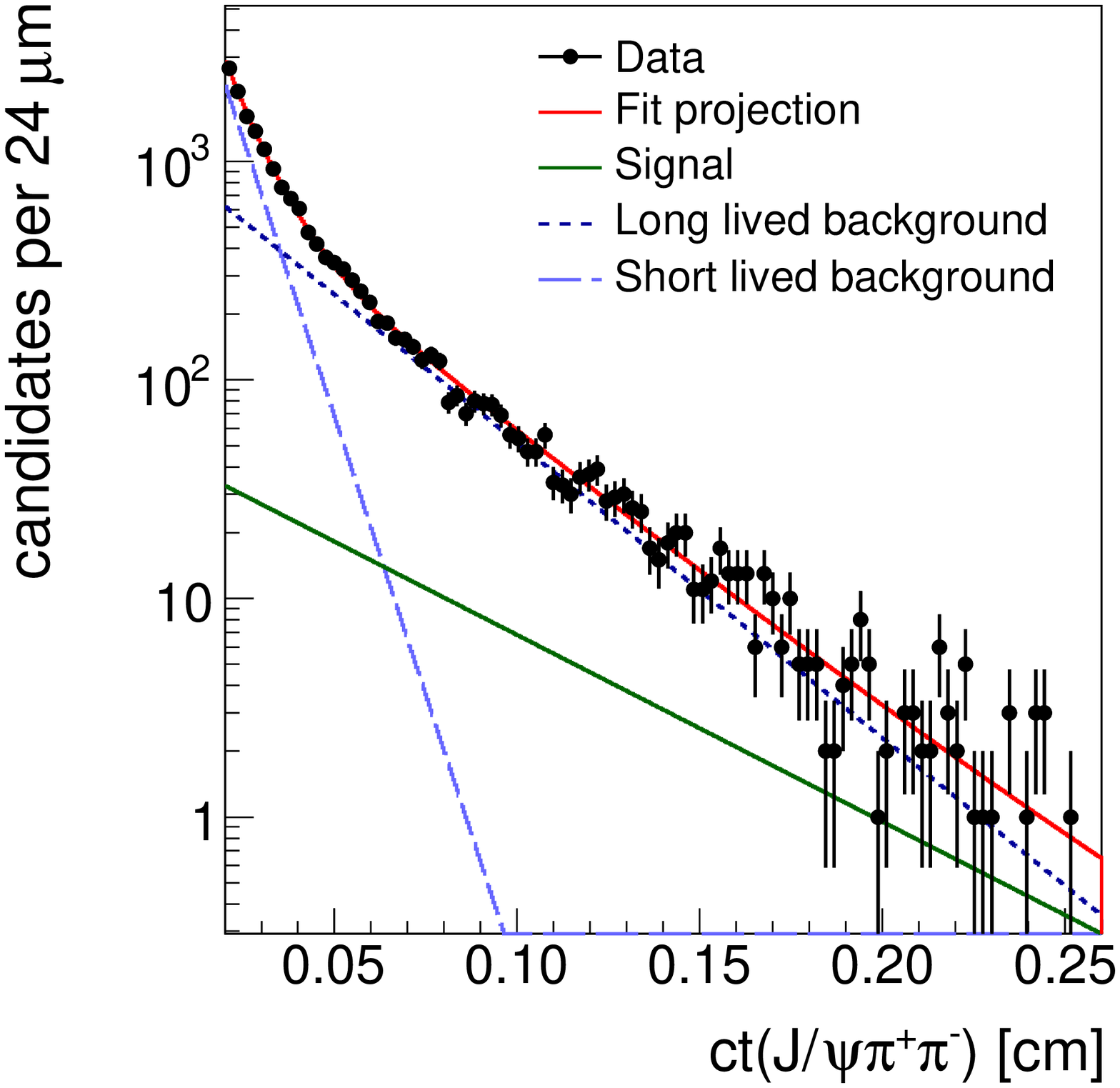}
\caption{(color online) Decay time distribution
with fit projection overlaid. 
}
\label{fig:defaultfit2}
\end{center}
\end{figure}
\begin{figure}
\begin{center}
\includegraphics[width=7.5cm]{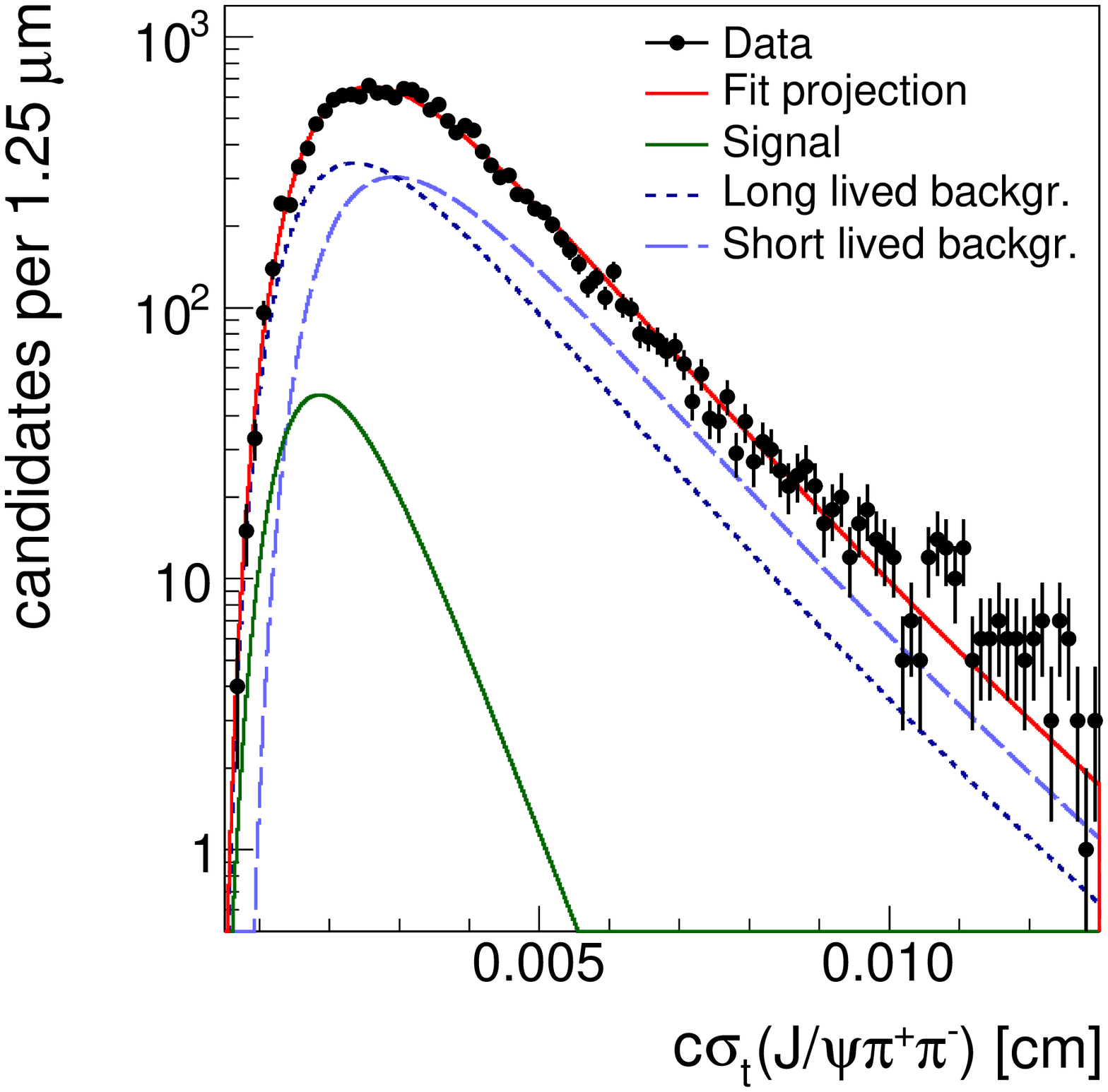}
\caption{(color online) Decay time uncertainty distribution
with fit projection overlaid. 
}
\label{fig:defaultfit3}
\end{center}
\end{figure}

Performing the likelihood fit to the selected data we extract
the \Bs lifetime in \BsJpsifzero decays
\begin{equation}
\tau(\BsJpsifzero) =
1.70_{-0.11}^{+0.12}(\mathrm{stat})\pm 0.03(\mathrm{syst})\,\mathrm{ps}.
\end{equation}
In Figs.~\ref{fig:defaultfit1} to \ref{fig:defaultfit3} we show the data together with the
projection of the fit.

\section{Conclusions}
\label{sec:conclusions}

We confirm the observation of the $\BsJpsifzero(980)$ decay from
the LHCb \cite{Aaij:2011fx} and Belle \cite{Li:2011pg} experiments. 
The observed signal is the world's largest and we perform the most precise
measurement of the ratio of branching fractions
$R_{f_0/\phi}$ between \BsJpsifzero and $\BsJpsiphi(980)$ decays:
\begin{eqnarray}
R_{f_0/\phi}&=&\frac{\mathcal{B}(\BsJpsifzero(980))}{\mathcal{B}(\BsJpsiphi)} 
\frac{\mathcal{B}(f_0(980)\rightarrow
\pi^+\pi^-)}{\mathcal{B}(\phi\rightarrow 
K^+K^-)}= \nonumber \\ && 0.257\pm0.020(\mathrm{stat})\pm0.014(\mathrm{syst}).
\end{eqnarray}
In this result we assume that the observed signal is solely due to the decay $\BsJpsifzero(980)$ and correct
for the acceptance of the invariant mass selection of the pion pair.
Using the world average \BsJpsiphi branching fraction
\cite{Nakamura:2010zzi} $R_{f_0/\phi}$ can be converted into the
product of branching fractions of
\begin{eqnarray}
\mathcal{B}(\BsJpsifzero(980))\mathcal{B}(f_0(980)\rightarrow
\pi^+\pi^-)= \nonumber \\ (1.63 \pm 0.12\pm 0.09 \pm 0.50 )\times 10^{-4},
\end{eqnarray}
where the first uncertainty is statistical, the second is systematic,
and the third one is due to the uncertainty on the 
\BsJpsiphi and $\phi \rightarrow K^+K^-$ branching fractions.
The measurement presented here agrees well with the previous
measurements of this quantity and with theoretical predictions.

Moreover, our sample allows us to
measure the \Bs lifetime in the $\BsJpsifzero(980)$ decay mode:
\begin{equation}
\tau(\BsJpsifzero(980)) =
1.70_{-0.11}^{+0.12}(\mathrm{stat})\pm 0.03(\mathrm{syst})\,\mathrm{ps}.
\end{equation}
This is the first measurement of the \Bs lifetime in a decay to a pure
\textit{CP} eigenstate. In the context of the standard model the
lifetime measured in this decay mode to a \textit{CP}-odd final state can be interpreted as
the lifetime of the heavy \Bs eigenstate. 
The measured value agrees well both
with the standard model expectation as well as with other
experimental determinations.

While the precision of the lifetime measurement is still limited by statistics,
it provides an important cross-check on the result determined
in \BsJpsiphi decays, which relies on an angular separation of
two \textit{CP} eigenstates.
Furthermore, the measured lifetime can be used as an external constraint in the \BsJpsiphi analysis 
to improve the determination of the \textit{CP}-violating phase in the \BsJpsiphi decay.
The lifetime measurement in $\BsJpsifzero(980)$ decays is also the next
step towards a tagged time dependent \textit{CP}-violation measurement, which 
can provide an independent constraint on the \textit{CP} violation in \Bs mixing.

\begin{acknowledgments}

\end{acknowledgments}
We thank the Fermilab staff and the technical staffs of the
participating institutions for their vital contributions.
This work was supported by the U.S. Department of Energy and
National Science Foundation; the Italian Istituto Nazionale
di Fisica Nucleare; the Ministry of Education, Culture,
Sports, Science and Technology of Japan; the Natural
Sciences and Engineering Research Council of Canada; the
National Science Council of the Republic of China; the Swiss
National Science Foundation; the A.P. Sloan Foundation; the
Bundesministerium f\"ur Bildung und Forschung, Germany; the
Korean World Class University Program, the National Research
Foundation of Korea; the Science and Technology Facilities
Council and the Royal Society, UK; the Institut National de
Physique Nucleaire et Physique des Particules/CNRS; the
Russian Foundation for Basic Research; the Ministerio de
Ciencia e Innovaci\'{o}n, and Programa Consolider-Ingenio
2010, Spain; the Slovak R\&D Agency; and the Academy of
Finland. 

%

\end{document}